# A New Method for Verifying $d$-MC Candidates

Wei-Chang Yeh
Department of Industrial Engineering and Engineering Management
National Tsing Hua University
P.O. Box 24-60, Hsinchu, Taiwan 300, R.O.C.
yeh@ieee.org

*Abstract* — Network reliability modeling and calculation is a very important study domain in reliability engineering. It is also a popular index for validating and measuring the performance of real-world multi-state flow networks (MFNs), e.g., the applications in internet of things, social networks, clouding computing, and 5G. The $d$-MC is a vector, the maximum flow of whose related network is $d$, and any vector less than the $d$-MC is not a $d$-MC in MFNs. The MFN reliability can be calculated in terms of $d$-MCs. Hence, the $d$-MC is one of the most popular tools for evaluating the MFN reliability. The method to find all $d$-MCs is through the mathematical programming whose solutions are called $d$-MC candidates, and all $d$-MCs are selected from these candidates. In this study, a novel and simple algorithm is proposed to filter out $d$-MCs from these $d$-MC candidates after removing duplicates. The time complexity of the proposed algorithm is analyzed along with the demonstration using an example. An experiment with 200 random networks is outlined to compare the proposed, traditional, and best-known algorithms used for verifying $d$-MC candidates.

**Acronym:**

MC         : Minimal cut.

$d$-MC     : $d$-Minimal cut.

MFN : Multistate Flow Network.

**Notations:**

|•|: the number of elements in •.

$Pr(•)$: the probability of event •.

$V$: $V=\{1, 2,…, n\}$ is the node set.

$E$: the arc set.

$W$: $W=(w_1, w_2,…, w_m)$ is the weight vector and $w_k$ is an non-negative integer for $k =1, 2, …, m$.

$G(V, E, W)$: a connected multi-state flow network with $V$, $E$, and $W$, where 1 and $n$ are the specified source node and sink node, respectively. For example, Fig. 1 is a connected multi-state flow network with $W = (3, 2, 2, 2, 2, 3)$.

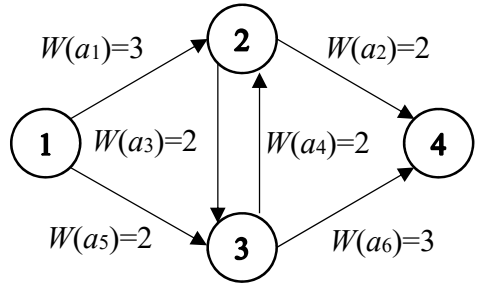

**Figure 1.** Example network

$a_k$: $a_k$ is the $k$th arc in $E$.

$e_{i,j}$: $e_{i,j} \in E$ is a directed arc from nodes $i$ to $j$.

$C_i$: the $i$th MC, e.g., $C_1 = \{a_1, a_5\}$ in Fig. 1.

$V(C)$: the MCV of MC $C$.

$E(V^*)$: $E(V^*) = \{e_{i,j} \in E \mid$ for all $i \in V^*$ and $j \notin V^*\}$, where $V^* \subseteq V$.

$d(C), d(C_i)$: the complete $d$-MC set generated from all MCs and MC $C_i$, respectively.

$d^{\#}(C), d^{\#}(C_i)$: the complete $d$-MC candidate set generated from all MCs and MC $C_i$, respectively.

$X_{i,j}$:   the $j$th $d$-MC candidate generated from $C_i$, e.g., $X_{1,1} = (3, 2, 2, 2, 0, 3)$ in Fig. 1.

$X(a_k)$:   the current capacity value of the $k$th arc in vector $X$, e.g., $X_{1,1}(a_1) = X_{1,1}(a_6) = 3$, $X_{1,1}(a_2) = X_{1,1}(a_3) = X_{1,1}(a_4) = 2$, and $X_{1,1}(a_5) = 0$ if $X_{1,1} = (3, 2, 2, 2, 0, 3)$.

$G(V, E, X_{i,j})$:   the subgraph induced from $G(V, E, W)$ by replacing $W$ with $X_{i,j}$. For example, Fig. 2 is $G(V, E, X_{2,2})$ obtained from Fig. 1, where $X_{2,2} = (3, 1, 2, 2, 2, 2)$.

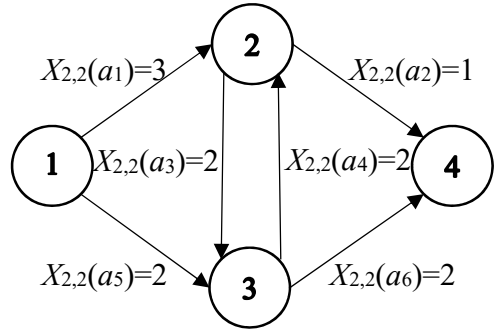

**Figure 2.** $G(V, E, X_{2,2})$ induced from Fig. 1.

$\Psi(G)$:   the state distribution of $G(V, E, W)$, e.g., Table 1 is the state distribution of Fig. 1.

**Table 1.** State distribution $\Psi$ of Fig. 1.

| $i$ | State | Probability | $i$ | State | Probability |
|---|---|---|---|---|---|
| 1 | 3 | 0.60 | 4 | 2 | 0.55 |
|   | 2 | 0.20 |   | 1 | 0.25 |
|   | 1 | 0.10 |   | 0 | 0.20 |
|   | 0 | 0.10 | 5 | 2 | 0.80 |
| 2 | 2 | 0.60 |   | 1 | 0.15 |
|   | 1 | 0.25 |   | 0 | 0.05 |
|   | 0 | 0.15 | 6 | 3 | 0.70 |
| 3 | 2 | 0.40 |   | 2 | 0.15 |
|   | 1 | 0.35 |   | 1 | 0.10 |
|   | 0 | 0.25 |   | 0 | 0.05 |

$F(X_{i,j})$:   the maximum flow in $G(V, E, X_{i,j})$, e.g., $F(X_{2,2}) = F((3, 1, 2, 2, 2, 2)) = 3$ in Fig. 2.

$F_d(X_{i,j})$:   the vector records one of the ways that the $d$ unit of flows is transmitted from nodes 1 to $n$ in $G(V, E, X_{i,j})$, e.g., $F_d(X_{2,2}) = (3, 1, 2, 0, 0, 2)$ in Fig. 2, where $X_{2,2} = (3, 1, 2, 2, 2, 2)$.

$R(V, E, X)$: the residual network after sending $d$ units of flow from nodes 1 to $n$ in $G(V, E, X)$, e.g., $X_{2,2} = (3, 1, 2, 2, 2, 2)$ and one of its residual networks $R(V, E, X_{2,2})$ after sending $d$ units of flow based on $F_d(X_{3,2}) = (3, 1, 2, 0, 0, 2)$ is shown in Fig. 3, where $d = 3$.

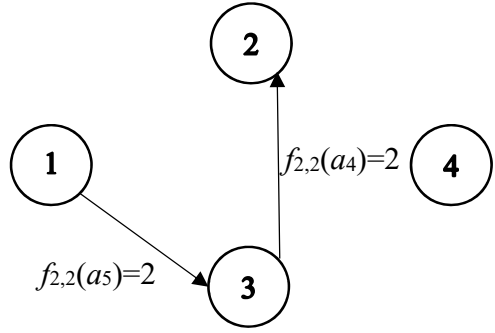

**Figure 3.** Residual network $R(V, E, X_{2,2})$ induced from Fig. 2.

$S(X)$: $S(X) = \{$node 1 and all the nodes which are reachable from node 1 in $R(V, E, X)\} \subseteq V$. For example, $S(X_{2,2}) = \{1, 2, 3\}$ in Fig. 3, where $X_{2,2} = (3, 1, 2, 2, 2, 2)$.

$T(X)$: $T(X) = \{$node $n$ and all the nodes which are reachable to node $n$ in $R(V, E, X)\} \subseteq V$. For example, $T(X_{2,2}) = T((3, 1, 2, 2, 2, 2)) = \{4\}$ in Fig. 3.

$o_j$: a vector such that $o_j(a_j)=1$ and $o_j(a_i)=0$ for all $i \neq j$.

$U(X)$: $U(X) = \{ a \in E \mid X(a) < W(a)\} \subseteq E$, e.g., $U(X_{2,2}) = \{a_2, a_6\}$ in Fig. 1, where $X_{2,2} = (3, 1, 2, 2, 2, 2)$.

$R_d$: the MFN reliability with level $d$ and it is the success probability that at least $d$ unit of flows is able to be sent from nodes 1 to $n$.

**Nomenclature:**

$d$-MC candidate: a system-state vector $X$ is a $d$-MC candidate if and only if $X$ satisfied the following equations [1-3]:

(1) $\sum_{\forall a \in C} X(a) = d$      (1)

(2) $X(a) \leq W(a)$, for all $a \in C$      (2)

(3) $X(a)=W(a)$, for all $a \notin C$   (3)

where $C$ is a MC.

$d$-MC: a system-state vector $X$ is a $d$-MC if and only if [1-3]

(1) $F(X)=d$;   (4)

(2) $F(X+o_j)>d$, where each $a_j \in E$ with $X(a_j)<W(a_j)$.   (5)

Unsaturated arc: $a_i$ is an unsaturated arc if and only if $a_i \in E$ with $X(a_i)<W(a_i)$

Duplicate remover: the method for removing duplicate $d$-MC candidates

$d$-MC filter: the method for filtering out real $d$-MCs from $d$-MC candidates

MCV: an MCV $V(C)$ is the node subset of the subgraph reachable from node 1 after removing the MC $C$.

$d$-MCV: the node subset $S(X)$ is called a $d$-MCV in $R(V, E, X)$.

**Assumptions:**

1) All nodes are 100% perfect.

2) The state of each arc is a non-negative integer in relation to a given distribution.

3) The states of different arcs are statistically independent.

4) The flow conservation law is obeyed.

5) The maximum flow is larger than $d$.

## 1. INTRODUCTION

A multistate flow network (MFN) is a special graph with multistate components representing different performance levels. It is possible to model various practical systems using MFNs, e.g., the internet of things [4], grid and cloud computing [5, 6], wireless sensor networks [7, 8], transportation systems [1], oil/gas production systems [1], power transmission and distribution systems [9-11], and Data Mining [12].

For example, a simple electric power transmission and distribution system is modeled as an MFN, as illustrated in Fig. 1. Here, $V = \{1, 2, 3, 4\}$, $E = \{a_1, a_2, a_3, a_4, a_5, a_6\}$, the source node is 1, the sink node is 4, the arc $a$ is marked by its maximal capacity $W(a)$, and the arc state distribution is listed in Table 1. This MFN is powered at different levels. Each arc, $a \in E$, represents a transmission line containing multiple physical transmission lines, such as T3 cable, E1 cable, and optical fiber. Each node, $v \in V$, represents a computer center containing multiple switches [15].

Network reliability is one of the most important factors in designing, evaluating, and validating network performance in our modern society [13]. However, it is an NP-hard problem to calculate the exact MFN reliability [3, 14]. Many algorithms have been proposed to calculate the MFN reliability [13-45]. Most exact algorithms are either cut-based [21, 37, 40-47] or path-based [15-20, 22, 23, 26-29, 34, 35, 36, 38, 39]. Both of these algorithm types are adapted from graph theory because the network originated from graph theory.

The MFN reliability is defined as the probability that the source node can send at least a predefined number, say $(d+1)$, of units of flow to a sink node to meet our requirement [1, 3, 14], that is [27]:

$R_{d+1} = 1 - \Pr(\{$ node 1 is able to send at most $d$ units of flow to node $n$ in $G(V, E, W)\})$ (6)

$= 1 - \Pr(\{X \mid F(X) \leq d \}).$ (7)

$= 1 - \Pr(X_1 \cup X_2 \cup \ldots \cup X_\chi)$, where $F(X_k) = d$, $\Pr(X_k) = \Pr(\{X \mid$ for all $X \leq X_k\})$, and $X_i \cup X_j = \{X \mid$ for all $X \leq X_i$ or $X \leq X_j\}$ (8)

From Eq. (7), all state vectors, say $X$, where the maximum flow in $G(V, E, X)$, i.e., $F(X)$, is less than or equal to $d$, are needed to calculate $R_{d+1}$. The state-space method [25] is based on Eq.

(7), and is required to find all such state vectors before calculating $R_{d+1}$. From Eq. (8), the MFN reliability $R_{d+1}$ can also be calculated by using state vectors $X_k$, where $X_k$ acts like the upper boundary points of level $d$, for $k = 1, 2, \ldots, \chi$, called $d$-MCs herein. Each $d$-MC is satisfied Eq. (7), i.e., the number of $d$-MCs is less than that of elements in Eq. (7). Each $d$-MC is satisfied when the number of $d$-MCs is less than that of elements in Eq. (7). Hence, it is more efficient to calculate $R_{d+1}$ based on Eq. (8), for which we need to find all $d$-MCs. Thus, $d$-MCs play an important role in the cut-based algorithms used for calculating the exact MFN reliability $R_{d+1}$ [21, 37, 40-42, 45].

The $d$-MC challenge is to search for all $d$-MCs, and a variety of methods have been proposed to solve this problem [37, 40-42, 45]. The $d$-MC methods are outlined in Section 2; most $d$-MC methods include four stages [37, 40-42, 45]:

1) Search for all MCs of which any of its proper subsets is not a real MC. For example, $\{a_2, a_3, a_5\}$ is an MC between nodes 1 and 4 in Fig. 1.

2) Generate all $d$-MC candidates, e.g., $X$, from each MC using Eqs. (1)-(3).

   For example, (3, $\underline{2}$, $\underline{1}$, 2, $\underline{0}$, 3), (3, $\underline{2}$, $\underline{0}$, 2, $\underline{1}$, 3), (3, $\underline{1}$, $\underline{2}$, 2, $\underline{0}$, 3), (3, $\underline{1}$, $\underline{1}$, 2, $\underline{1}$, 3), (3, $\underline{1}$, $\underline{0}$, 2, $\underline{2}$, 3), (3, $\underline{0}$, $\underline{2}$, 2, $\underline{1}$, 3), and (3, $\underline{0}$, $\underline{1}$, 2, $\underline{2}$, 3) are all 3-MC candidates generated from MC $\{a_2, a_3, a_5\}$ using Eqs. (1)-(3) in Fig. 1. Note that to easily understand the relationship between the MC, the $d$-MC, and its candidates, the arc positions marked with the underscore in each vector belong to the same MC. For example, (3, $\underline{2}$, $\underline{1}$, 2, $\underline{0}$, 3) is generated from the MC $\{a_2, a_3, a_5\}$ as the second, third, and fifth coordinates are underscored.

3) Filter out real $d$-MCs and remove duplicates.

   For example, the 3-MC candidates generated from MC $\{a_2, a_3, a_5\}$ listed above are not all real 3-MCs, and some must be eliminated to increase the efficiency of calculating $R_{d+1}$ using $d$-MCs (stage 4). Additionally, two different MCs may generate the same $d$-MC. Hence, even though these $d$-MC candidates are real $d$-MCs, there are some duplicates that should be eliminated before calculating the MFN reliability. For example, the real 3-MC $X = (0, \underline{2}, 1, \underline{2}, \underline{2}, \underline{2})$ is generated from two different MCs $\{a_1, a_3, a_6\}$ and $\{a_1, a_2, a_3\}$ in Fig. 4.

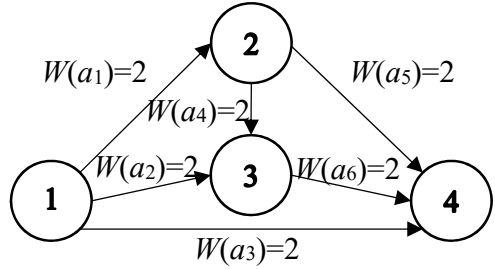

**Figure 4.** Example of duplicate $d$-MCs.

4) Calculate the MFN reliability, $R_{d+1}$, in terms of the identified $d$-MCs using Eq. (8), the inclusion–exclusion method [50, 51], and the sum-of-disjoint product method [18, 48, 49].

The four stages above have been studied extensively and are all *NP*-Hard problems [37, 40-42, 45]. However, in stage 3, there is still room to improve the efficiency of filtering out the real $d$-MCs from the $d$-MC candidates obtained in stage 2.

The value of the final MFN reliability $R_{d+1}$ is unchanged with duplicate $d$-MCs or when including $d$-MC candidates that are not real (as calculated in Eq. (8)). However, the run time doubles if there is one duplicate or one $d$-MC candidate that is not real [37, 40-42, 45]. This is a characteristic of the NP-hard problem. Hence, the run time will increase up to $2^\delta$ times if the total number of $d$-MC candidates and duplicates is $\delta$. Therefore, the purpose of this study is to develop a more intuitive, efficient, and novel algorithm than the current, best-known $d$-MC algorithm for filtering out the real $d$-MC from $d$-MC candidates and to remove all duplicate $d$-MCs.

The rest of the paper is organized as follows. Section 2 provides a summary of the existing $d$-MC algorithms including the best-known algorithm, integration of the unsaturated arc test and the $d$-MC-to-MC comparison method, and the traditional candidate-to-candidate comparison method. Section 3 describes the novel $d$-MCV concept which is the fundamental of the proposed algorithm, the proposed new $d$-MC filter, and the new duplicate remover method. Section 4 describes the pseudo code of the proposed algorithm in detail and removing duplicate $d$-MC candidates and filtering out real $d$-MCs, along with a discussion of its accuracy and complexity. In Section 5, the proposed algorithm is exemplified by a step-by-step example to demonstrate how to filter out real $d$-MCs from $d$-MC candidates in an MFN. Another 200 randomly generated networks from $n = 10, 20, \ldots, 100$ are tested to compare the efficiency of the proposed, best-known, and traditional algorithms for filtering out real $d$-MCs. Section 6 provides the concluding remarks.

## 2. CURRENT ALGORITHMS FOR THE $d$-MC PROBLEM

Each real $d$-MC is a $d$-MC candidate meeting the requirements of Eqs.(1)–(3) [37, 40-42, 45]. However, not all $d$-MC candidates are real $d$-MCs [37, 40-42, 45], as discussed in the third part of Section 1. Moreover, a real $d$-MC may be generated from different MCs. Therefore, there is a need to filter out real $d$-MCs from $d$-MC candidates and remove duplicate $d$-MCs. Previously, there were two primary methods for this: 1) the candidate-to-candidate comparison method and 2) the unsaturated arc test method in filtering out $d$-MCs and the $d$-MC-to-MC comparison method for removing duplicates [37, 40-42, 45]. The details of these two methods are discussed below:

### 2.1 Candidate-to-Candidate Comparison Method [53]

The candidate-to-candidate comparison method compares each $d$-MC candidate to others that have not been removed yet. To illustrate this method, we use Eq. (8), where $d$-MC is the upper-bound vector for calculating the MFN reliability $R_{d+1}$ such that [37, 40-42, 45, 53]:

1) A state vector less than any $d$-MC is already included in the unions of $d$-MCs, as shown in Eq. (8). Hence, it is unnecessary to find vectors that are less than any $d$-MC.

2) If state vector $X$ is larger than any $d$-MC, of which the maximum flow is equal to $d$, its maximum flow $F(X)$ is larger than $d$ in $G(V, E, X^*)$, and therefore it is unnecessary to identify this kind of vector.

Therefore, $d$-MC candidate $X$ is not a real $d$-MC and can be removed immediately if $X$ is less than any other $d$-MC candidate. The time complexity of the candidate-to-candidate comparison method is $O(m \cdot |d^{\#}(C)|^2)$ [37, 40-42, 45]. It is calculated from the required number of comparisons, where $d^{\#}(C)$ is the complete $d$-MC candidate set and $|d^{\#}(C)|$ is the number of $d$-MC candidates in $d^{\#}(C)$.

After removing these 3-MC candidates $X_{i,j}$ with $F(X_{i,j}) < d$, the candidate-to-candidate comparison method is implemented on the remaining candidates in the column under "C2C" of Table 2. For example, $X_{i,j} < X_{1,1}$ in the row for $i = 3$ and $j = 1$ means that $X_{3,1} < X_{1,1}$ and $X_{i,j}$ is not a real 3-MC.

**Table 2.** Two major ways to complete stage 3.

| $i$ | $j$ | $C_i$ | $X_{i,j}$ | $F(X_{i,j})$ | C2C[1] | $U_{Arc}$[2] | $S(X_{i,j})$ | $T(X_{i,j})$ |
|---|---|---|---|---|---|---|---|---|
| 1 | 1 | $\{a_1, a_5\}$ | ($\underline{3}$, 2, 2, 2, $\underline{0}$, 3) | 3 | | | {1} | {2, 3, 4} |
| | 2 | | ($\underline{2}$, 2, 2, 2, $\underline{1}$, 3) | 3 | | | {1} | {2, 3, 4} |
| | 3 | | ($\underline{1}$, 2, 2, 2, $\underline{2}$, 3) | 3 | | | {1} | {2, 3, 4} |
| 2 | 1 | $\{a_2, a_6\}$ | (3, $\underline{2}$, 2, 2, 2, $\underline{1}$) | 3 | | | {1, 2, 3} | {4} |
| | 2 | | (3, $\underline{1}$, 2, 2, 2, $\underline{2}$) | 3 | | | {1, 2, 3} | {4} |
| | 3 | | (3, $\underline{0}$, 2, 2, 2, $\underline{3}$) | 3 | | | {1, 2, 3} | {4} |
| 3 | 1 | $\{a_2, a_3, a_5\}$ | (3, $\underline{2}$, $\underline{1}$, 2, $\underline{0}$, 3) | 3 | $X_{i,j} < X_{1,1}$ | $P(X_{i,j}+o_3) = 0$ | {1}* | |
| | 2 | | (3, $\underline{2}$, $\underline{0}$, 2, $\underline{1}$, 3) | 3 | | | {1, 2} | {3, 4} |
| | 3 | | (3, $\underline{1}$, $\underline{2}$, 2, $\underline{0}$, 3) | 3 | $X_{i,j} < X_{1,1}$ | $P(X_{i,j}+o_2) = 0$ | {1}* | |
| | 4 | | (3, $\underline{1}$, $\underline{1}$, 2, $\underline{1}$, 3) | 3 | | | {1, 2} | {3, 4} |
| | 5 | | (3, $\underline{1}$, $\underline{0}$, 2, $\underline{2}$, 3) | 3 | | | {1, 2} | {3, 4} |
| | 6 | | (3, $\underline{0}$, $\underline{2}$, 2, $\underline{1}$, 3) | 3 | $X_{i,j} < X_{2,3}$ | $P(X_{i,j}+o_5) = 0$ | {1, 2} | {4}# |
| | 7 | | (3, $\underline{0}$, $\underline{1}$, 2, $\underline{2}$, 3) | 3 | $X_{i,j} < X_{2,3}$ | $P(X_{i,j}+o_3) = 0$ | {1, 2} | {4}# |
| 4 | 1 | $\{a_1, a_4, a_6\}$ | ($\underline{3}$, 2, 2, $\underline{0}$, 2, $\underline{0}$) | 2 | | | | |
| | 2 | | ($\underline{2}$, 2, 2, $\underline{1}$, 2, $\underline{0}$) | 2 | | | | |
| | 3 | | ($\underline{2}$, 2, 2, $\underline{0}$, 2, $\underline{1}$) | 3 | $X_{i,j} < X_{2,1}$ | $P(X_{i,j}+o_1) = P(X_{i,j}+o_4) = 0$ | {1, 3} | {4} |
| | 4 | | ($\underline{1}$, 2, 2, $\underline{2}$, 2, $\underline{0}$) | 2 | | | | |
| | 5 | | ($\underline{1}$, 2, 2, $\underline{1}$, 2, $\underline{1}$) | 3 | $X_{i,j} < X_{2,1}$ | $P(X_{i,j}+o_1) = P(X_{i,j}+o_4) = 0$ | {1}* | |
| | 6 | | ($\underline{1}$, 2, 2, $\underline{0}$, 2, $\underline{2}$) | 3 | $X_{i,j} < X_{1,3}$ | $P(X_{i,j}+o_4) = P(X_{i,j}+o_6) = 0$ | {1}* | |
| | 7 | | ($\underline{0}$, 2, 2, $\underline{2}$, 2, $\underline{1}$) | 2 | | | | |
| | 8 | | ($\underline{0}$, 2, 2, $\underline{1}$, 2, $\underline{2}$) | 2 | | | | |
| | 9 | | ($\underline{0}$, 2, 2, $\underline{0}$, 2, $\underline{3}$) | 2 | | | | |

[1] The candidate-to-candidate comparison method.
[2] The unsaturated arc test and any one item in $U_{Arc}$ is not held, the related $d$-MC candidate is not a real $d$-MC.
* Not a $d$-MC from Theorem 1 discussed in Section 3.4.
# Not a $d$-MC from Lemma 3 discussed in Section 3.4.

**2.2** Unsaturated Arc Test Method with the $d$-MC-to-MC Comparison Method

The unsaturated arc test method first proposed by Yeh in [40] remains the best-known algorithm for filtering out real $d$-MCs from $d$-MC candidates. To illustrate this method, we assume $X$ is a $d$-MC candidate where $F(X) = d$. Using the definition of the MFN reliability $R_{d+1}$, we only need to test $F(X+o_j) > d$ for each unsaturated arc $a_i \in E$, if we want to verify whether $X$ is a real $d$-MC candidate. This concept can be used for time complexity $[O(m) O(n) + O(md)] O(|d^\#(C)|) = O(md |d^\#(C)|)$ [40]. Here, $O(m)$ is the maximal number of unsaturated arcs for each $d$-MC candidate, $O(md)$ is the time complexity to calculate $F(X)$ using the Ford–Fulkerson algorithm [52], and $O(n)$ is the time complexity to calculate $P(X+o_j)$, i.e., find a path from nodes 1 to $n$ in $G(V, E, X+o_j)$ [41].

For example, refer to the unsaturated arc test demonstrated in the column under "U$_{arc}$" of Table 2. In the row for $i = 4$ and $j = 3$, $P(X_{i,j}+o_1) = P(X_{i,j}+o_4) = 0$ indicates that there is no path from nodes 1 to 4 in both $G(V, E, X_{4,3}+o_1)$ and $G(V, E, X_{4,3}+o_4)$. Therefore, $X_{4,3}$ is not a real 3-MC and must be discarded. Note that a $d$-MC candidate $X_{i,j}$ is not a real $d$-MC if $P(X_{i,j}+o_k) = 0$ for at least one $k$ with $X_{i,j}(a_k) < W(a_k)$.

Yeh's unsaturated arc test is more efficient than the traditional candidate-to-candidate comparison method discussed above [40, 41]. However, the unsaturated arc test can only filter out real $d$-MCs from $d$-MC candidates, it cannot remove duplicates. Therefore, it is necessary to solve the problem of duplicate $d$-MCs by comparing each $d$-MC to all other $d$-MCs. This is called the $d$-MC to $d$-MC comparison test. To improve the time taken to remove duplicate $d$-MCs without using the $d$-MC to $d$-MC comparison test, Yeh proposes that the $d$-MC-to-MC comparison method should be used after each $d$-MC is verified to be a real $d$-MC. The time complexity of the $d$-MC-to-MC comparison method is $O(m) \cdot O(|C|) \cdot O(|d^{\#}(C)|) = O(m \cdot |C| \cdot |d^{\#}(C)|)$ [40, 41]. From the above, the time complexity of the unsaturated arc test method and the $d$-MC-to-MC comparison method is $O(md\, |d^{\#}(C)|) + O(m \cdot |C| \cdot |d^{\#}(C)|) \approx O(m \cdot |C| \cdot |d^{\#}(C)|)$ [40, 41]. This is less time than that of the traditional candidate-to-candidate comparison test given $|C| << |d^{\#}(C)|$.

## 3. NOVEL $d$-MCV, NEW DUPLICATE REMOVER, AND $d$-MC FILTER

A new concept called the $d$-MCV, extended from the MCV proposed in [44], is proposed to filter out $d$-MCs from $d$-MC candidates, remove duplicates among $d$-MCs, and to reduce run time.

## 3.1 Novel $d$-MCV

This section explains how the new $d$-MCV is calculated.

The MCV $V(C_i)$ is the node subset of the subgraph reachable from node 1 after removing the MC $C_i$. The arc subset $C_i$ is an MC only if the node subset $V(C_i)$ is an MCV. For example, let MC $C_1 = \{a_1, a_3, a_6\}$ and MC $C_2 = \{a_1, a_2, a_3\}$, as shown in Fig. 4; we have $V(C_1) = \{1, 3\}$ and $V(C_2) = \{1\}$.

The node subset $S(X)$ is called a $d$-MCV of $d$-MC candidate $X$ in the residual network $R(V, E, X)$ after send $d$ units of flow from nodes 1 to $n$ in $G(V, E, X)$. The $d$-MCV is obtained based on the residual network. The residual network can be easily determined by any algorithm in finding the maxmum flow, e.g., the Ford–Fulkerson algorithm [52], and the time complexity to find a residual network is at most equal to that of any maxmum flow algorithm.

In the Ford-Fulkerson algorithm [52], if there is a rule to adhere to when selecting a path then it would result in the selection of a distinct, unique augmented path. This would mean that there would only be one residual network. An example of a rule would be to always select the arc with the largest label among all unselected arc that has the same capacity with the previous node, say v, in the current augmented path from node 1 to $v$.

For example, $X = (0, \underline{2}, 1, \underline{2}, \underline{2}, \underline{2})$ is a 3-MC in Fig. 4 and $G(V, E, X)$ is shown in Fig. 5(a). In Fig. 5(a), the first augmented path starts from node 1 and is connected to node 3, but not node 4 as $X(a_1) = \text{Max}\{X(e_{1,2}) = X(a_1) = 0, X(e_{1,3}) = X(a_2) = 2, X(e_{1,4}) = X(a_3) = 1\}$ (as shown in Fig. 5(b)). Note that, the labels of $e_{1,3}$ and $e_{1,4}$ are 2 and 3, respectively. The first augmented path needs to be amended to connect to node 4 but not node 3 if $X(e_{1,4}) = X(a_3) = 1$ is changed to 2.

As shown in Fig. 5(b), $F_d(X) = (0, 2, 1, 0, 0, 2)$ describes the way based on the rule mentioned in the above to send $d$ units of flow from nodes 1 to $n$ in $G(V, E, X)$ and the residual subgraph $R(V,$ $E$, $X$) is shown in Fig. 5(c). The nodes reachable from node 1 in $R(V, E, X)$ are called the $d$-MCV $S(X)$, e.g., {1} in Fig. 5(c).

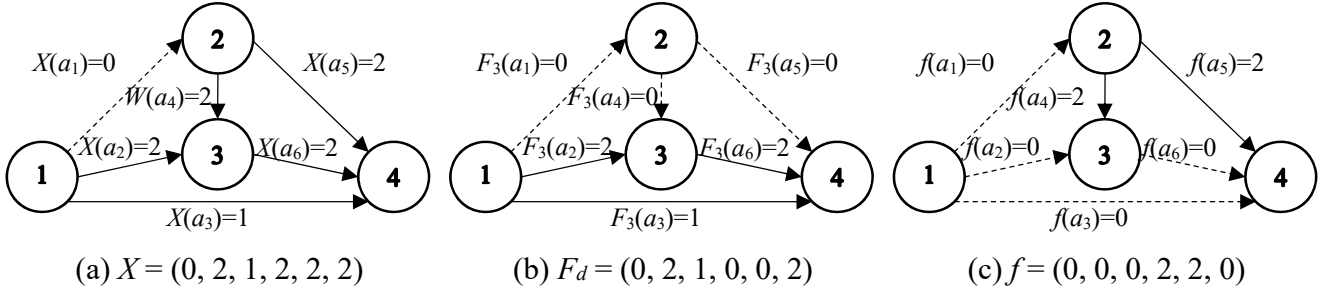

(a) $X = (0, 2, 1, 2, 2, 2)$  (b) $F_d = (0, 2, 1, 0, 0, 2)$  (c) $f = (0, 0, 0, 2, 2, 0)$
**Figure 5.** $G(V, E, X)$, $G(V, E, F_d(X))$, and $R(V, E, X) = G(V, E, f)$.

The following lemma discusses that there is only one $d$-MCV regardless of the related $d$-MC candidates generated from the many different MCs.

**Lemma 1.** Let $d$-MC $X_i$ and $X_j$ be generated from different MCs. $S(X_i) = S(X_j)$ if $d$-MCs $X_i = X_j$.

**Proof.** Because there is only one residual network and each residual network also has only one $S(X)$, there is one and only one $d$-MCV of a $d$-MC no matter which MC generates it. □

### 3.2 New $d$-MC Filter

The new $d$-MC filter can be explained in terms of the concept of the $d$-MCV as follows.

From the definition of the $d$-MC, a $d$-MC candidate $X$ generated from MC $C$ with $F(X) = d$ is a real $d$-MC if $F(X+o_j) > d$, where each $a_j \in E$ with $X(a_j) < W(a_j)$. From the definition of the $d$-MC candidate, $X$ is generated from an MC, say $C$, and $C(a) \leq W(a)$ for all $a \in C$, and $C(a) = W(a)$ for all $a \notin C$. Hence, in the definition of the $d$-MC, "$a_j \in E$" can be simplified to "$a_j \in C$" as only the states of these arcs in $C$ are less than or equal to their maximal states.

Also, in the definition of the $d$-MC, say $X$, $F(X+o_j) > d$ is equivalent to the direct path from nodes 1 to $n$ in $R(V, E, X+o_j)$ via $a_j$ for all $a_j \in C$ with $X(a_j) < W(a_j)$. The state of arcs in MC $C$ generated $X$ are all zeros in $R(V, E, X)$. Hence, $F(X+o_j) > d$ is also identical to the direct path from nodes 1 to $n$, after increasing the state of only arc (say $a_j \in C$ with $X(a_j) < W(a_j)$) from zero to one in $R(V, E, X)$. Thus, we have the following equivalent statements:

1) $F(X+o_j) > d$;

2) there is a direct path from nodes 1 to $n$ in $R(V, E, X+o_j)$ via $a_j$;

3) there is a direct path from nodes 1 to $n$ after increasing the state of one arc in $U(X)$ from zero to one in $R(V, E, X)$.

(9)

After removing MC $C$, we have $i \in V(C)$, and $j \notin V(C)$ for all $e_{i,j} \in C$. Let $X$ be a $d$-MC candidate. Similarly, the residual network $R(V, E, X)$ is separated into at least two connected subgraphs, formed by $S(X)$ and $T(X)$. The other subgraphs cannot go to node $n$ nor are they reachable from node 1 if $S(X) \cup T(X) \neq V$. The following lemma shows the first rule based on the novel $d$-MCV to verify $d$-MC candidates.

**Lemma 2.** If $|S(X) \cup T(X)| = n$, $d$-MC candidate $X$ is a real $d$-MC.

**Proof.** Because $|S(X) \cup T(X)| = n$, we have $S(X) \cup T(X) = V$. Moreover, node 1 is connected to any node in $S(X)$ and any node not in $S(X)$ is connected to node $n$, i.e., there is a path from nodes 1 to $n$ via arc $a_k$ in $R(V, E, X+o_k)$ for all $a_k$ with $X(a_k) < W(a_k)$. Hence, $F(X + o_k) > F(X) = d$ and $X$ is a real $d$-MC. □

For example, in Fig. 1, $X_{3,2} = (3, \underline{2}, \underline{0}, 2, \underline{1}, 3)$ is a 3-MC candidate generated from $C_3 = \{a_2 (= e_{2,4}), a_3 (= e_{2,3}), a_5 (= e_{2,4})\}$. The $G(V, E, X_{3,2})$ and $R(V, E, X)$ are shown in Fig. 6(a) and 6(b), where $F_d(X) = (2, 2, 0, 0, 1, 1)$, respectively. As shown in Fig. 6(b), $S(X) = \{1, 2\}$, $T(X) = \{3, 4\}$, and $S(X) \cup T(X) = V$. Hence, $X_{3,2}$ is a real 3-MC. Note that $a_4 \notin U(X_{3,2})$ and $a_4 \notin C_3$.

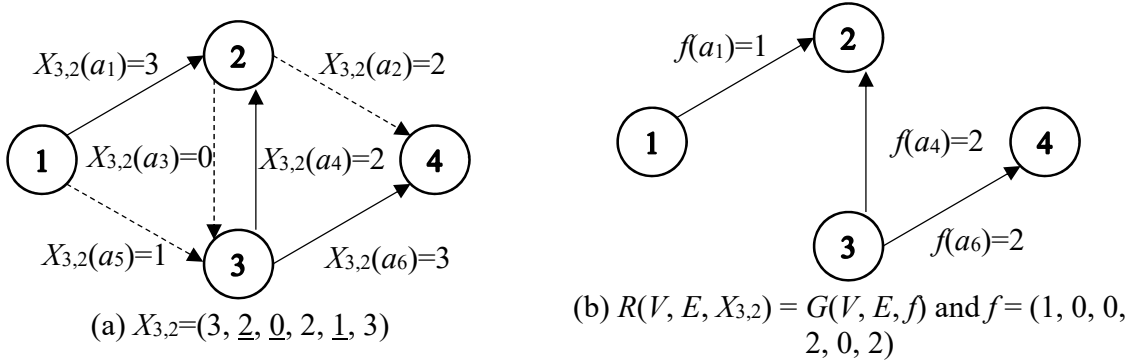

(a) $X_{3,2} = (3, \underline{2}, \underline{0}, 2, \underline{1}, 3)$

(b) $R(V, E, X_{3,2}) = G(V, E, f)$ and $f = (1, 0, 0, 2, 0, 2)$

**Figure 6.** The example network explains Lemma 2.

The following lemma shows the second rule also the last one based on the novel $d$-MCV to verify $d$-MC candidates.

**Lemma 3.** Assume that $|S(X) \cup T(X)| < n$. If $X(e_{i,j}) < W(e_{i,j})$ for at least one $a_k = e_{i,j}$ with $i \in S(X)$ and $j \notin (S(X) \cup T(X))$, or with $i \notin (S(X) \cup T(X))$ and $j \in T(X)$, then $X$ is not a real $d$-MC.

**Proof.** Let $X^* = X + o_k$, $a_k = e_{i,j}$, $X(e_{i,j}) < W(e_{i,j})$, $i \in S(X)$, and $j \notin (S(X) \cup T(X))$. Because $j \notin (S(X) \cup T(X))$, there is no path from nodes $j$ to $n$ in $R(V, E, X+o_k)$ and $F(X^*) = F(X + o_k) = F(X) = d$ from Eq. (5). In the same way, we have the same result for $i \notin (S(X) \cup T(X))$ and $j \in T(X)$. Note that there is no need to test any saturated arc $a$, i.e., $X(a) = W(a)$, from the definition of the $d$-MC. Hence, this lemma is true. □

For another example, the 3-MC candidate $X_{3,1} = (3, \underline{2}, \underline{1}, 2, \underline{0}, 3)$ is generated from MC $C_3 = \{a_2, a_3, a_5\}$ in Fig. 1. In the residual network of $X_{3,1}$ shown in Fig 7(b), we have $S(X_{3,1}) = \{1\}$, $T(X_{3,1}) = \{3, 4\}$, and $|S(X_{3,1}) \cup T(X_{3,1})| < |V| = 4$. Also, $a_3 = e_{2,3}$, $X(a_3) = 1 < W(a_3) = 2$, $2 \notin (S(X_{3,1})$

∪ $T(X_{3,1})$) ={1, 3, 4}, and 3 ∈ $T(X_{3,1})$ = {3, 4}. Hence, increasing the state of $a_3$ from 0 to 1 in Fig. 7(b) only identifies a path from nodes 2 to 3, but not from nodes 1 to 4. Thus, $X_{3,1}$ is not a real 3-MC.

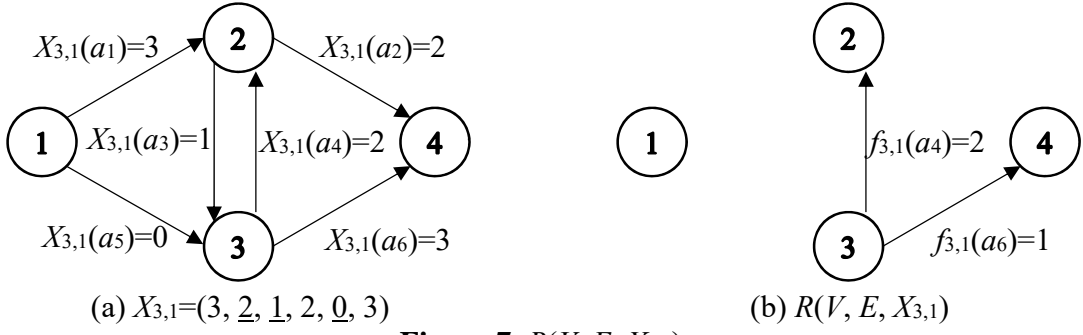

(a) $X_{3,1}$=(3, 2, 1, 2, 0, 3)  (b) $R(V, E, X_{3,1})$

**Figure 7.** $R(V, E, X_{3,1})$.

Lemmas 2 and 3 both take only $O(n)$ to scan $R(V, E, X)$ to verify whether the $d$-MC candidate $X$ is a real $d$-MC based on the proposed saturated boundary.

### 3.3 New $d$-MC Duplicate Remover

The $d$-MCV $S(X)$ is also the key to verifying whether the $d$-MC candidate $X$ is a duplicate by using $E(S(X))$. Note that, like the traditional candidate-to-candidate method, the proposed algorithm verifies whether the $d$-MC candidate is a duplicate first, discards it if it is a duplicate, or verifies it further to check if it is a real $d$-MC. The proposed $d$-MC duplicate remover method is discussed next.

The following lemma discusses some paramount roles of $E(S(X))$ for $d$-MC $X$.

**Lemma 4.** $E(S(X))$ is an MC and minimum cut in $G(V, E, W)$ and $G(V, E, X)$, respectively, for all $d$-MC $X$.

**Proof.** $G(V, E, W)$ has no path from any nodes in $S(X)$ to any nodes that are not in $S(X)$ after removing $E(S(X))$, i.e., $E(S(X))$ is a cut in $G(V, E, W)$. If $E(S(X)) - \{e_{a,b}\}$ is still a cut in $G(V, E, W)$ for all $e_{a,b} \in E(S(X))$, we have either node $a \notin S(X)$ or $b \in S(X)$ and both are impossible. Hence, no arc is redundant in $E(S(X))$, i.e., $E(S(X))$ is also a MC in $G(V, E, W)$. Moreover, because $X$ is a $d$-MC and there is no path from nodes 1 to $n$ in $R_d(V, E, X)$, the maximum flow in $G(V, E, X)$ is $d$ and $\sum_{e \in E(S(X))} X(e) = d$, i.e., $E(S(X))$ is a minimum cut in $G(V, E, X)$. Hence, this lemma is correct. □

The following lemma shows the relationship between $d$-MC $X$ and $E(S(X))$.

**Lemma 5.** The $d$-MC $X$ generated from MC $C$ in $G(V, E, W)$ is also a $d$-MC generated from $E(S(X))$ in $G(V, E, W)$.

**Proof.** From Lemma 4, $E(S(X))$ is a MC in $G(V, E, W)$. If $a_j \notin E(S(X))$ with $X(a_j) < W(a_j)$, then $F(X + o_j) = F(X) = d$ because $\sum_{\forall e \in E(S(X))} X(e) = d$ and $a_j \notin E(S(X))$, i.e., $X$ is not a $d$-MC which is contradictory to our assumption. Hence, $X(e) \leq W(e)$ for all $e \in E(S(X))$ and $X(e) = W(e)$ for all $e \notin E(S(X))$, i.e., $X$ is also generated from MC $E(S(X))$ from Eqs. (1) - (3). □

The following lemma is one of the major contributions of the paper. It shows that we can discard the $d$-MC candidate or $d$-MC in which its $d$-MCV is not equal to the MCV of the MC that generated it.

**Theorem 1.** Let $d$-MCs $X$ and $X_i$ be generated from MCs $C$ and $C_i$, respectively, $C_i \neq C$, $C_i \neq C_j$, and $X = X_i$ for all $i, j \in I$. There is one and only one $k \in I$ such that $S(X) = V(C_k)$, i.e., $S(X) \neq V(C_i)$ for all $i \in I - \{k\}$.

**Proof.** From Lemma 1, we have $S(X) = S(X_i)$ for all $i \in I$. From Lemma 5, at least $X$ is also a $d$-MC generated from MC $C^*$, i.e., $S(X) = V(C_k)$ for some $C_k = C^*$. Moreover, $V(C^*) \neq V(C^\#)$ if and only if $C^* \neq C^\#$ from the property of the MCV [44]. Hence, it is impossible to have another duplication, say $X_l$ generated from $C_l$ such that $S(X) = V(C_k) = V(C_l)$ if $l \neq k$. From the above, this lemma is true. □

From Theorem 1, $d$-MC candidate $X$ generated from MC $C$ with $F(X) = d$ can be removed without losing any real $d$-MC if $S(X) \neq V(C)$.

Referring to the same example discussed in Section 3.1, the real 3-MC $X = (0, \underline{2}, 1, \underline{2}, \underline{2}, \underline{2})$ was generated from two different MCs $C_1 = \{a_1, a_3, a_6\}$ and $C_2 = \{a_1, a_2, a_3\}$ in Fig. 4. However, only the one generated from $C_2$ was retained, and the one generated from $C_1$ was discarded because $S(X) = \{1\} = V(C_2)$. Also, only one $X$ is saved in the end, i.e., there are no duplicates and no lost real $d$-MCs.

It takes only $O(|V(C)|) \ll O(n)$ to scan the subgraph formed by $V(C)$ in $R(V, E, X)$ to decide whether $X$ is a duplicate $d$-MC candidate by checking $S(X) = V(C)$ in Theorem 1.

### 3.4. PROPOSED ALGORITHM AND TIME COMPLEXITY

The pseudo code of the proposed algorithm is outlined here to filter out real $d$-MCs, without duplicates, from $d$-MC candidates based on the novel $d$-MCV.

**Input:** An MFN $G(V, E, W)$ with a source node 1, a sink node $n$, the MC set, and the $d$-MC candidate set generated from each MC.

**Output:** The complete $d$-MC set $d(C)$ in $G(V, E, W)$.

**STEP 0.** Let $i = j = 1$ and $C_d = \emptyset$.

**STEP 1.** If $F(X_{i,j}) < d$, $X_{i,j}$ is not a real $d$-MC and go to STEP 6.

**STEP 2.** If $S(X_{i,j}) \neq V(C_i)$, $X_{i,j}$ is either not a real $d$-MC or a duplicate based on Theorem 1 and go to STEP 6

**STEP 3.** If $|S(X_{i,j}) \cup T(X_{i,j})| = n$, go to STEP 5 based on Lemma 2.

**STEP 4.** Implement Lemma 3 to verify whether $X$ is a real $d$-MC, Go to STEP 6 if $X$ is not a real $d$-MC.

**STEP 5.** Let $C_d = C_d \cup \{X_{i,j}\}$.

**STEP 6.** If $j < |d^\#(C_i)|$, let $j = j + 1$ and go to STEP 1.

**STEP 7.** If $i = |C|$, halt and $C_d$ is the complete $d$-MC set $d(C)$. Otherwise, let $i = i + 1, j = 1$, and go to STEP 1.

The basic concepts behind the proposed algorithm are Lemmas 2, 3 and Theorem 1 implemented by scanning the graph for each $d$-MC candidate with time complexity $O(n)$ after finding the maximum flow. The summary of the time complexity necessary for verifying one $d$-MC candidate and all $d$-MC candidates are listed in Table 3 after finding the maximum flow. Note that $O(m)=O(n^2)$.

Table 3. Time complexities for all algorithms.

| Algorithms | Verify one $d$-MC candidate without considering duplicates | Verify one $d$-MC candidate with considering duplicates | Total Time Complexity |
|---|---|---|---|
| Alg[1] | $O(n)$ | $O(n)$ | $O(n\,|d^\#(C)|)$ |
| U$_{Arc}$[2] | $O(md + mn)$ | $O(md + mn + m|C|)$ | $O(m|C|\,|d^\#(C)|)$ |
| C2C[3] | $O(md + m\cdot|d^\#(C_d)|)$ | $O(md + m\cdot|d^\#(C_d)|)$ | $O(m\cdot|d^\#(C_d)|^2)$ |

[1] The proposed algorithm.
[2] The unsaturated arc test and the $d$-MC-to-MC comparison method.
[3] The candidate-to-candidate comparison method.

Thus, the time complexity of the proposed algorithm is faster than the existing ones. It is faster for verifying a $d$-MC candidate and/or all $d$-MC candidates with and/or without considering the duplications, as shown in Table 3.

## 4. STEP-BY-STEP EXAMPLE AND COMPUTATIONAL EXPERIMENTS

The proposed $d$-MC algorithm is demonstrated by a step-by-step example together with its performance test in this section.

**4.1** Step-by-step example in verifying $d$-MC candidates

For convenience and ease of understanding the proposed algorithm, the example network is shown in Fig. 1, where nodes 1 and 4 are the source and sink nodes, respectively. It is illustrated to demonstrate the general procedure of the proposed algorithm, step-by-step, for verifying the $d$-MCs of the network reliability algorithm.

**Solution:**

**STEP 0.** Let $i = j = 1$ and $C_d = \emptyset$.

**STEP 1.** Because $F(X_{1,1}) = 3$, go to STEP 2.

**STEP 2.** Because $S(X_{1,1}) = V(C_1) = \{1\}$, go to STEP 3.

**STEP 3.** Because $|S(X_{1,1}) \cup T(X_{1,1})| = n = 4$, where $T(X_{1,1}) = \{2, 3, 4\}$, $X_{1,1}$ is a real $d$-MC without duplicate and go to STEP 5.

**STEP 5.** Let $C_d = C_d \cup \{X_{1,1}\} = \{(\underline{3}, 2, 2, 2, \underline{0}, 3)\}$.

**STEP 6.** Because $j = 1 < |d^{\#}(C_1)| = 3$, go to STEP 1.

:

**STEP 6.** Because $j = |d(C_2)| = 3$, go to STEP 7.

**STEP 7.** Because $i = 2 < |C| = 4$, let $i = i + 1 = 3, j = 1$, and go to STEP 1.

**STEP 1.** Because $F(X_{3,1}) = d = 3$, go to STEP 2.

**STEP 2.** Because $S(X_{3,1}) = \{1\} \neq V(C_3) = \{1, 2\}$, go to STEP 6.

$$\vdots$$

**STEP 1.** Because $F(X_{4,1}) < 3$, discard $X_{4,1}$ and go to STEP 6.

$$\vdots$$

**STEP 1.** Because $F(X_{4,3}) = d = 3$, go to STEP 2.

**STEP 2.** Because $S(X_{4,3}) = V(C_4) = \{1, 3\}$, go to STEP 3.

**STEP 3.** Because $|S(X_{4,3}) \cup T(X_{4,3})| = 3 < n = 4$, where $T(X_{4,3}) = \{4\}$, go to STEP 4.

**STEP 4.** (Lemma 3) Because $F(X_{4,1}) = F(X_{4,1} + o_3) = 3$, $X_{4,1}$ is not a real 3-MC and go to STEP 6.

$$\vdots$$

The final result of filtering out all $d$-MCs, without duplicates, from $d$-MC candidates using the proposed algorithm is listed in the last columns of Table 2. From Table 2, we have nine 3-MCs:

$$c_1 = (3, 2, 2, 2, 0, 3), c_2 = (2, 2, 2, 2, 1, 3), c_3 = (1, 2, 2, 2, 2, 3),$$

$$c_4 = (3, 2, 2, 2, 2, 1), c_5 = (3, 1, 2, 2, 2, 2), c_6 = (3, 0, 2, 2, 2, 3),$$

$$c_7 = (3, 2, 0, 2, 1, 3), c_8 = (3, 1, 1, 2, 1, 3), c_9 = (3, 1, 0, 2, 2, 3).$$

After using the inclusion-exclusion method [50, 51] and the arc states provided in Table 2, we have

$$R_4 = 1 - Pr(c_1 \cup c_2 \cup \ldots \cup c_9) = 1 - 0.5488750109 = 0.451124989.$$

**4.2 Computational experiments**

From the time complexity discussed in Section 4, the proposed algorithm is more efficient than the existing algorithms in verifying $d$-MC candidates. The performance of the proposed algorithm was further tested on 200 distinctive MFNs to demonstrate that it is faster.

Each network structure of the 200 MFNs was generated randomly from $n = 10, 20, 30, \ldots, 100$, and the number of arcs was also generated randomly in the experiments. Furthermore, $d = W(a) =$ (the minimum of the degree of nodes 1 and $n$) for all arcs $a$ was generated randomly and tested 20 times, i.e., 200 MFNs.

When verifying $d$-MC candidates, the proposed algorithm (named Alg), the existing best-known algorithm (named $U_{Arc}$) [40, 41], and the traditional $d$-MC to $d$-MC pairwise comparison algorithm (named C2C) [40], were implemented in C/C++ programming language with the time limit is 5 hours. For a fair comparison, these three algorithms were coded, tested, and run on an Intel Core i7 3.07 GHz PC with 32 GB memory.

**Table 4.** Average runtimes for finding $d$-MCs.

| $n$ | Alg | $U_{Arc}$ | C2C |
|---|---|---|---|
| 10 | 6.29323E-07 | 6.31179E-06 | 0.002055934 |
| 20 | 8.0599E-06 | 0.00028123 | 1.072606217 |
| 30 | 0.000142001 | 0.00426209 | 135.4220568 |
| 40 | 0.001927224 | 0.106412049 | 2948.184184 |
| 50 | 0.031567277 | 1.869965108 | * |
| 60 | 0.487015617 | 58.19042296 | * |
| 70 | 6.786522465 | 507.3575246 | * |
| 80 | 82.95454318 | 12632.75546 | * |
| 90 | 1362.024746 | * | * |
| 100 | 21231.46971 | * | * |

* the program is forced to terminate

The experimental results are shown in Table 4; the average runtime for each network size from $n = 10, 20, \ldots, 100$ are reported. As shown in Table 4, we can observe that the run time increases exponentially with the network size for all algorithms. This is due to it being NP-hard to filter out all $d$-MCs from $d$-MC candidates [1, 40, 41].

Alg is the fastest algorithm and C2C is the slowest, which was expected from the theoretical results based on time complexity. Both Alg and $U_{Arc}$ can verify whether a $d$-MC candidate is a real $d$-MC based on its own vector without needing to compare with others, which is the basis of the candidate-to-candidate comparison method. Hence, these two methods are more efficient than C2C. Also, Alg can remove duplicates based on its own vector (residual network); however, $U_{Arc}$ can only achieve that by comparing all found $d$-MCs with all MCs. Thus, Alg is much faster than $U_{Arc}$.

The program is forced to terminate for C2C and $U_{Arc}$ if $n \geq 40$ and $n \geq 80$, respectively. The reason for the above phenomenon is that $d = W(a)$ for all arcs $a$ in the experiment and the number of $d$-MCs generated by MC $C_i$ is bounded by $|C_i| \leq \min\{ \binom{|C_i|+d-1}{d},$ $\prod_{\forall a_j \in C_i|}[W(a_j)+1] = (d+1)^{|C_i|} \} = \delta_i$ and $|C_i| << \delta_i$ [40, 41]. Thus, the proposed algorithm, Alg, outperforms the traditional algorithms, $U_{Arc}$ and C2C, for all $n$. Note that the explanation can also be observed from the time complexity.

## 5. CONCLUSIONS

Various innovative and important technologies, such as the internet, 5G, clouding/edge/fog computing, and block chain, are all built on network models. Network reliability is a general tool for authenticating, designing, and evaluating the performance of network models. Hence, it is always important to improve its efficiency when calculating the networks.

The $d$-MC is a popular method for calculating the MFN reliability, and all $d$-MCs are filtered out from $d$-MC candidates before calculating the MFN reliability $R_{d+1}$ in terms of real $d$-MCs. An efficient, simple, and novel $d$-MC filter was developed for filtering out the real $d$-MCs after finding all the $d$-MC candidates. By simply exploiting the structure of residual network after sending $d$ units of flow in the $d$-MC candidate related sub-network, the accuracy and the time complexity of the proposed algorithm was relatively easy to prove and analyze.

Regarding time complexity, the proposed $d$-MC filter outperforms the existing algorithm. There is a noteworthy improvement in the time complexity $O(n)$ over the previous $O(md)$ in verifying one $d$-MC candidate, where $O(m)=O(n^2)$.

From an extensive experimental study on 200 benchmark networks, the proposed algorithm clearly outperformed the best-known unsaturated arc test and the conventional candidate-to-candidate comparison method. Hence, the proposed $d$-MC filter is very useful in improving the efficiency of verifying $d$-MC candidates from both theoretical and practical aspects.

## ACKNOWLEDGMENT

We wish to thank the anonymous editor and the referees for their constructive comments and recommendations, which significantly improved this paper. his research was supported in part by the Ministry of Science and Technology of Taiwan, R.O.C. under grant MOST 107-2221-E-007-072-MY3.